\documentclass[a4paper,10pt]{edpsci}
\usepackage{amsmath}
\usepackage{amssymb}
\usepackage{graphicx}
\usepackage{epstopdf}
\usepackage[numbers]{natbib}
\usepackage{hyperref}
\usepackage{url}
\graphicspath{{./figs/}}

\newcommand{\beginsupplement}{%
        \setcounter{section}{0}
        \renewcommand{\thesection}{S\arabic{section}}%
        \setcounter{table}{0}
        \renewcommand{\thetable}{S\arabic{table}}%
        \setcounter{figure}{0}
        \renewcommand{\thefigure}{S\arabic{figure}}%
     }

\newcommand{\dir}{\boldsymbol{\varphi}}
\newcommand{\tar}{\boldsymbol{t}}

\hypersetup{colorlinks=true,citecolor=blue,urlcolor=blue,linkcolor=blue}

\usepackage[ruled]{algorithm2e}

\begin{document}
\journalname{Acta Acustica}
\articletype{Technical and Applied Paper}

\title{Statistical validation and full-sphere extension of a Bayesian model for human static sound localisation}

\titlerunning{Likelihood-based validation of a Bayesian sound localisation model}



\author{Roberto Barumerli\inst{1}\correspondingauthor{}
\and
Fabian Brinkmann\inst{2}
\and
Emanuele Zanoni\inst{3}
\and
Anton Hoyer\inst{2}
\and
Lorenzo Picinali\inst{1}
\and
Michele Geronazzo\inst{3,1}
}

\authorrunning{Barumerli et al.}


\institute{Dyson School of Design Engineering, Imperial College London, London, United Kingdom
\and
Audio Communication Group, Technische Universit\"{a}t Berlin, Germany
\and
Department of Industrial Systems Technology and Management, University of Padova, Vicenza, Italy}

\abstract{
Auditory models provide a computational framework for understanding the mechanisms enabling human spatial hearing. However, their validation typically relies on performance metrics rather than likelihood-based fitting and formal model comparison. Here, we present two contributions building on the Bayesian sound localisation model (Barumerli et al., 2023), which jointly infers horizontal and vertical sound directions by comparing noisy perceptual features against templates derived from individual head-related transfer functions (HRTFs). First, we formulate an explicit likelihood function for directional responses in static conditions and validate it through a complete Bayesian workflow. Parameter recovery on simulated responses confirms that free parameters are identifiable, though recovery degrades for the spectral and prior parameters at extreme values, and fits to behavioural data from 33 participants (collated across three sound localisation experiments sharing a common condition with individual HRTFs) yield convergent, interpretable parameter estimates. Second, we apply this framework to resolve an open question in the model's formulation: whether the interpolation method used to interpolate templates onto a quasi-uniform spherical grid affects localisation predictions. Comparing the original interpolation against three alternatives, spatial coverage across the full sphere and preservation of spectral content at high frequencies are the primary determinants of template quality, while the specific interpolation algorithm is secondary once these conditions are met. Together, these contributions demonstrate how established model-based statistical methods can strengthen both fundamental spatial hearing research and applied problems such as HRTF evaluation. To facilitate reproducible research and integration with modern machine learning workflows, this work releases an open-source Python implementation.}



\maketitle

\section{Introduction}
Auditory computational models are essential tools for understanding the perceptual mechanisms underlying spatial hearing, formalising hypotheses about how the auditory system extracts and employs spatial information~\cite{dau_auditory_2008, meddis_computational_2010}. Yet, the statistical methods used to calibrate and evaluate models vary widely across studies, often relying on heuristic metric matching that provides no principled basis for goodness-of-fit assessment or model comparison~\cite{barumerli_frambi_2025}. In computational neuroscience, this problem has been addressed through a shared framework combining probabilistic model implementation with formal statistical analysis, where Bayesian inference provides a common language for building models and testing hypotheses against experimental data~\cite{kriegeskorte_cognitive_2018}. Psychoacoustics has yet to broadly adopt this methodology, and in this work, we demonstrate that such a formal statistical workflow can be successfully applied to an auditory model for sound localisation.
To address this methodological gap, we recently introduced \textsc{FrAMBI}, a software framework for auditory modelling based on Bayesian inference~\cite{barumerli_frambi_2025}. \textsc{FrAMBI} provides a standardised structure for implementing auditory models as probabilistic processes, enabling likelihood computation for parameter estimation and Bayesian model comparison. This approach offers a fundamental advantage over behavioural metrics traditionally used for model evaluation~\cite{carlile_nature_1997, dietz_framework_2018}: such metrics aggregate responses into scalar averages, providing descriptive summaries rather than a direct statistical measure of model fit~\cite{pitt_when_2002}. A probabilistic formulation, in contrast, enables likelihood-based parameter estimation: the likelihood measures how probable the observed responses are under the model, given a specific parameter configuration~\cite{myung_tutorial_2003}. This is, to our knowledge, the only framework that simultaneously allows parameter recovery validation and hypothesis testing through model comparison. Parameter recovery is crucial: if parameters cannot be recovered from simulated data, the model is misspecified, and its predictions carry no inferential validity, regardless of how plausible they appear~\cite{wilson_ten_2019}. Moreover, model comparison quantifies which of two competing model variants better accounts for the observed data~\cite{kass_bayes_1995}. \textsc{FrAMBI} demonstrated these capabilities on proof-of-concept localisation models, and its application to models with complex, non-linear processing stages has so far been limited (e.g.,~\cite{llado_spectral_2025}).

Sound source localisation is a central paradigm in the study of human spatial hearing, and computational models are the primary tools for formalising and testing hypotheses about its underlying mechanisms~\cite{majdak_amt_2022}. These models range from binaural processing of interaural cues for horizontal localisation~\cite{dietz_framework_2018} to spectral cue evaluation against learned templates for the vertical dimension~\cite{baumgartner_modeling_2014}, and full probabilistic integration of both~\cite{barumerli_bayesian_2023}. Beyond fundamental research, the Barumerli et al.\cite{barumerli_bayesian_2023} model enabled perception-based evaluation of non-individual head-related transfer function (HRTF) selection~\cite{daugintis_classifying_2023, daugintis_perceptual_2026}, HRTF measurement consistency across laboratories~\cite{daugintis_listener_2025, bahu_toward_2026}, HRTF upsampling~\cite{hogg_hrtf_2024,hogg_listener_2025}, evaluation of numerically generated HRTFs~\cite{meyer_accuracy_2025}, and prediction of localisation impairment due to head-worn devices~\cite{llado_predicting_2024}. Yet the absence of a unified statistical methodology limits reproducibility and cross-study comparisons, and the Barumerli et al. model is an ideal candidate to address this gap: despite its fully probabilistic formulation, its original parameter estimation relied on an ad hoc metric-matching procedure rather than the likelihood implied by the model itself~\cite{myung_tutorial_2003}.

In this technical paper, we present two contributions building on the Barumerli et al.\ model~\cite{barumerli_bayesian_2023}. The first is methodological: we formulate an explicit likelihood function, validate it through parameter recovery on simulated data, and fit it to behavioural localisation responses from 33 participants, applying the statistical framework introduced in~\textsc{FrAMBI}. The second is applied: we use this framework to resolve an open question in the model's original formulation concerning how measured HRTFs are interpolated onto a spherical grid to form the internal templates used for direction inference. By comparing the original interpolation method against three full-sphere alternatives, we illustrate how principled statistical inference enables hypothesis testing that aggregate performance metrics cannot support. Together, these two contributions demonstrate that the same statistical machinery serves both ends of the research spectrum: on the fundamental side, it exposes structural model properties and identifiability limits that are invisible to metric-matching parameter estimation; on the applied side, the validated likelihood offers a behaviourally grounded metric for HRTFs in machine learning pipelines~\cite{hogg_listener_2025}. Finally, an open-source Python implementation is released alongside this work to facilitate reproducible research and integration with modern signal processing workflows.

\section{Materials and Methods}

This section is organised as follows. We first describe the model and how its likelihood function is derived. We then present the likelihood-fitting procedure, validated on simulated data, before applying it to behavioural localisation responses from 33 participants. Finally, we perform hypothesis testing by comparing four model variants, each using a different template interpolation method, to identify which best accounts for the observed localisation behaviour.

\subsection{Model formulation in brief}\label{sec:model}

Following~\cite{barumerli_bayesian_2023}, the model simulates static sound localisation as a Bayesian inference process~\cite{ma_bayesian_2023}: the listener extracts noisy spatial features, infers the most probable source direction, and maps it to a pointing response. The following details how the listener response for a single localisation trial is modelled.

The model starts by extracting a set of noisy spatial features $\mathbf{t}$ from a binaural stimulus generated by a single sound source with unknown direction $\dir$ and flat spectrum. The flat-spectrum assumption ensures that spatial features are determined solely by the listener's HRTF rather than by source spectral content:
\begin{equation}
    \tar = [x_{\text{itd}}, x_{\text{ild}}, \boldsymbol{x}_{\text{L,mon}}, \boldsymbol{x}_{\text{R,mon}}] + \boldsymbol{\delta}.
    \label{eq:features}
\end{equation}

The spatial cues are: interaural time difference $x_{\text{itd}}$ (ITD), interaural level difference $x_{\text{ild}}$ (ILD), and monaural spectral features for the left and right ears.  The ITD is estimated via interaural cross-correlation on the low-pass filtered binaural signal ($<3$\,kHz) and transformed following~\cite{reijniers_ideal-observer_2014} to obtain a dimensionless quantity with approximately constant variance across the lateral range, effectively compensating for the direction-dependent increase in ITD discrimination thresholds~\cite{mossop_lateralization_1998}. The ILD is approximated as the time-averaged broadband level difference between the two channels~\cite{andreopoulou_identification_2017}. Further, monaural cues were specified either as spectral amplitudes (with dimension $28\times1$)~\cite{middlebrooks_sound_2015} or positive spectral gradients (with dimension $27\times1$)~\cite{baumgartner_modeling_2014}, both derived from a gammatone filterbank with $N_B=28$ non-overlapping equivalent rectangular bandwidth bands spanning [0.7, 18] kHz; gradients are differences between adjacent bands, yielding one fewer dimension than amplitudes. Further details on cue computation are available in~\cite{barumerli_bayesian_2023}.
Further, $\boldsymbol{\delta}$ in Eq.\ref{eq:features} accounts for the uncertainties introduced by the hearing system by adding Gaussian noise $\mathcal{N}(0, \boldsymbol{\Sigma})$ with zero mean and diagonal covariance matrix $\boldsymbol{\Sigma}$ defined as:
\begin{equation}
    \label{eq:sigma}
    \boldsymbol{\Sigma} =\left[
    \begin{array}{ccc}
        \sigma_\mathrm{itd}^2  & 0 & 0 \\
        0 & \sigma_\mathrm{ild}^2 & 0 \\
        0 & 0 & \sigma_\mathrm{mon}^2 \boldsymbol{I}
    \end{array}\right],
\end{equation}
with $\sigma_\mathrm{itd}^2$ and $\sigma_\mathrm{ild}^2$ being the variances associated with the ITDs and ILDs and $\sigma_\mathrm{mon}^2\boldsymbol{I}$ being the covariance matrix for the monaural features where the scalar $\sigma_\mathrm{mon}$ represents a constant and identical uncertainty for all frequency bands and $\boldsymbol{I}$ is the identity matrix of size $(2\cdot28 \times 2\cdot28)$ if using spectral amplitudes and $(2\cdot27 \times 2\cdot27)$ if using spectral gradients.

The auditory pathway decodes spatial information from such cues to extract the direction of the incoming sound, assuming a fixed listening distance, $\dir = (\alpha, \beta)$ with $\alpha$ being the lateral or horizontal angle and $\beta$ being the rising or vertical angle within the horizontal-polar reference system~\cite{morimoto_localization_1984, middlebrooks_virtual_1999}. This reference system will be used to compute lateral and vertical errors. Bayes' law describes the inference of sound direction at the computational level~\cite{ma_bayesian_2023}, and assumes that the listener computes the posterior probability distribution over all possible sound directions~$\dir$ from the spatial features in~$\tar$ by combining the sensory likelihood $p(\tar | \dir)$ with prior beliefs about the sound direction $p(\dir)$:
\begin{equation}
    \label{eq:bayes}
    p(\dir | \tar) \propto p(\tar | \dir)p(\dir).
\end{equation}

The sensory likelihood function relies on the comparison of observed features~$\tar$ with templates~$\boldsymbol{s}(\dir)$ that contain reference features (i.e. noiseless) of Eq.~\ref{eq:features} for every sound direction~${\dir}$ and $\boldsymbol{\delta}=0$~\cite{middlebrooks_sound_2015,barumerli_bayesian_2023}. Our implementation estimates such templates as noiseless spatial cues computed from the listener-specific HRTFs sampled on a quasi-uniform spherical grid (see Sec.\ref{ssec:interpolation}). Therefore, the sensory likelihood function is:
\begin{equation}
\label{eq:sensory_evidence}
    p(\tar | \dir) = p(\tar | \boldsymbol{s}(\dir))=\mathcal{N}(\tar|\boldsymbol{s}(\dir), \boldsymbol{\Sigma}),
\end{equation}
where $\boldsymbol{\Sigma}$ corresponds to the uncertainties introduced in the observed features following the assumption that the auditory system has learned them through experience~\cite{ma_bayesian_2023}.

The a-priori probability~$p(\dir)$ reflects long-term expectations of listeners and is modelled as uniform along the horizontal dimension, consistent with evidence that azimuth responses are accurate and unbiased across listeners~\cite{ege_accuracy-precision_2018}. Further, the prior is centred towards the horizon in the vertical dimension, consistent with behavioural evidence that listeners bias responses towards eye level~\cite{ege_accuracy-precision_2018}:
\begin{equation}
        p(\dir) \propto \text{exp}\left(-\frac{\epsilon^2}{2\sigma^2_{prior}}\right),
    \label{eq:prior}
\end{equation}
with~$\epsilon$ denoting the vertical angle of~$\dir$ in spherical coordinates and~$\sigma_\mathrm{prior}^2$ the variance of the prior distribution~\cite{barumerli_bayesian_2023}. Importantly, $\epsilon$ is the elevation in standard spherical coordinates, equal to $\beta$ only in the median plane, so that the prior favours near-horizontal directions independently of the horizontal angle.

From the posterior distribution (Eq.~\ref{eq:bayes}), the model estimates the listener-internal sound direction by selecting the most probable direction of the source~$\dir$ by employing the maximum a-posteriori (MAP) estimate (i.e., direction with highest posterior probability):
\begin{equation}
    \label{eq:map}
    \hat{{\dir}}' = \arg\max_{{\dir}}  p(\tar | \boldsymbol{s}(\dir)) p({\dir}).
\end{equation}

Finally, the listener's response $\dir$ accounts for the post-decision process of transforming the internal estimation to a pointing response by adding direction-independent noise, modelled as isotropic following prior work~\cite{barumerli_bayesian_2023, baumgartner_modeling_2014}:
\begin{equation}
    \label{eq:response}
    \boldsymbol{\hat{\varphi}} = \boldsymbol{\hat{\varphi}}' + \boldsymbol{m},
\end{equation}
where~$\boldsymbol{m}$ is drawn from a von Mises-Fisher distribution (i.e., the spherical analogue of a multivariate Gaussian) with zero mean and concentration parameter~$\kappa_m$ (i.e., $\text{vMF}(0, \kappa_m)$)~\cite{mardia2009directional}. While isotropy is a simplification~\cite{bahu_comparison_2016}, we discuss the consequences of this assumption in Sec.~\ref{sssec:postchecks}. For reporting clarity, we report the motor noise as a standard deviation~$\sigma_{\text{m}}$ in degrees, converted to the von Mises-Fisher concentration~$\kappa_m$ via the relation~$\bar{R} = \exp(-\sigma_{\text{m}}^2/2)$, where~$\bar{R} = I_1(\kappa_m)/I_0(\kappa_m)$ is the mean resultant length (i.e. the expected length of the average unit response vector, with 0 indicating uniform dispersion and 1 perfect concentration) with modified Bessel functions of the first kind of order 0 and 1 ($I_0$ and $I_1$, respectively)~\cite{mardia2009directional}.

\begin{figure}
    \centering
    \includegraphics[width=0.9\linewidth]{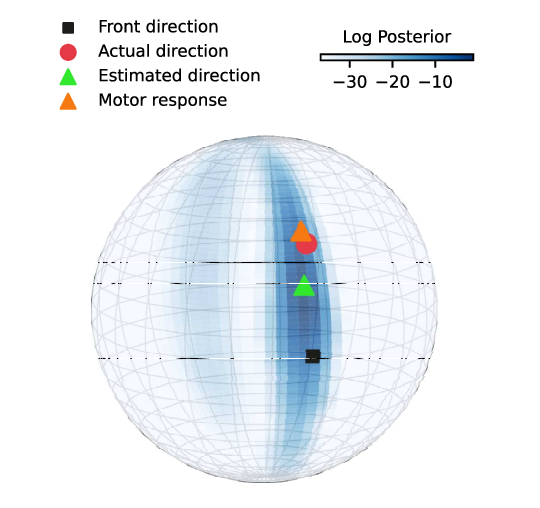}
    \caption{Illustration of a single modelled localisation trial for a sound presented at $\dir$ (red dot). Bluescale map: log-posterior distribution computed by evaluating Eq.~\ref{eq:bayes} for all directions of the template grid. Red dot: actual source direction $\dir$. Green triangle: listener-internal model estimate (Eq.~\ref{eq:map}). Orange triangle: final listener response (Eq.~\ref{eq:response}). Front direction is indicated with a black square for reference.}
    \label{fig:posterior}
\end{figure}

Figure~\ref{fig:posterior} reports a visual example of this process for a single estimation. By repeating this process across many repetitions and source positions, the model accounts for variability across trials and generates response distributions~\cite{barumerli_frambi_2025} from which localisation metrics such as lateral error, polar error, and quadrant error rate can be computed, following the same analytical approach as in localisation experiments~\cite{carlile_nature_1997, middlebrooks_virtual_1999,  majdak_3-d_2010}.

\subsection{Parameter estimation and model validation}

We derive an explicit likelihood function from the model's responses to enable principled parameter estimation and statistical validation~\cite{wilson_ten_2019, barumerli_frambi_2025}.

\subsubsection{Model-Likelihood Function}

In a localisation experiment with $N$ trials, where a sound is presented at direction $\dir_n$ and the listener responds with $\hat{\dir}^*_n$, the likelihood measures the model's ability to generate the observed responses given its parameters $\boldsymbol{\theta} = \{\sigma_\mathrm{itd}, \sigma_\mathrm{ild}, \sigma_\mathrm{mon}, \sigma_\mathrm{prior}, \kappa_m\}$~\cite{ma_bayesian_2023}:
\begin{multline}
    \mathcal{L}\!\left(\boldsymbol{\theta} \mid \left[\hat{\dir}^*_1, \ldots, \hat{\dir}^*_N\right], \left[\dir_1, \ldots, \dir_N\right]\right) = 
    \\ \prod_{n=1}^{N} p\!\left(\hat{\dir}^*_n \mid \dir_n, \boldsymbol{\theta}\right).
    \label{eq:loglik}
\end{multline}

Computing the likelihood requires marginalising over the model's uncertainties: perceptual uncertainty, capturing variability in the listener's internal direction estimate due to sensory noise and prior beliefs; and motor uncertainty, capturing variability in translating that estimate into a pointing response. For a sound presented at direction $\dir_n$, the likelihood marginalises over model-generated responses $\hat{\dir}$:
\begin{multline}
    p(\hat{\dir}^*_n \,|\, \dir_n, \boldsymbol{\theta}) = \\
    \int 
    p(\hat{\dir}' \,|\, \dir_n, \sigma_{\text{prior}}, \boldsymbol{\Sigma}) \, 
    p(\hat{\dir}^*_n \,|\, \hat{\dir}', \kappa_m) \, 
    d\hat{\dir}',
    \label{eq:lik_integral}
\end{multline}
where the two terms within the integral correspond to the perceptual stage (Eq.~\ref{eq:map}) and the motor stage (Eq.~\ref{eq:response}), respectively.

Since the integral in Eq.~\ref{eq:lik_integral} does not have an analytical solution, we resort to Monte Carlo integration, which provides an approximated solution~\cite{bishop_pattern_2006}. For each target direction $\dir_n$, we generate $M$ internal estimates $\{\hat{\dir}'\}_{m=1}^M$ using ancestral sampling~\cite{barumerli_frambi_2025}: each sample is drawn from the perceptual stage~(Eq.~\ref{eq:map}) by perturbing the spatial features with sensory noise and decoding the MAP estimate. Then, the motor stage~(Eq.~\ref{eq:response}) is applied analytically through the von Mises–Fisher kernel:
\begin{equation}
    p(\hat{\dir}^*_n \,|\, \dir_n, \theta) \approx \frac{1}{M} \sum_{m=1}^{M} 
    \text{vMF}(\hat{\dir}^*_n \,|\, \hat{\dir}'_m, \kappa_m).
    \label{eq:lik_mc}
\end{equation}
where Monte Carlo samples $M$ is selected based on the likelihood stability analysis in Sec.~\ref{ssec:param_id}.

Parameter estimates are then obtained by maximising the likelihood function:
\begin{equation}
    \boldsymbol{\hat \theta} = \underset{\boldsymbol\theta} {\mathrm{argmax}}~\mathcal{L}\!\left(\boldsymbol{\theta} \mid \left[\hat{\dir}^*_1, \ldots, \hat{\dir}^*_N\right], \left[\dir_1, \ldots, \dir_N\right]\right).
\end{equation}

\subsubsection{Likelihood estimation}
\label{sec:fitting}
The parameter vector $\boldsymbol{\theta}$ contains five parameters that jointly determine the model's predictions across both horizontal and vertical directions. However, not all parameters are identifiable from behavioural data: parameters that contribute additively to the response variance along the same spatial dimension cannot be separated, as any combination yielding the same total variance produces identical response distributions. Within our settings, $\kappa_m$ contributes to response variance in both horizontal and vertical dimensions, and it is confounded with $\sigma_\mathrm{itd}$ and $\sigma_\mathrm{ild}$ in the horizontal dimension and with $\sigma_\mathrm{mon}$ in the vertical dimension.

Fixing nuisance parameters (i.e., those that are not of primary interest and vary little across individuals) to theoretically motivated values is preferable to estimating them freely, and preserves the interpretability of the remaining parameters~\cite{pitt_when_2002}.
To do so, we exploit the structure of the horizontal-polar coordinate system~\cite{morimoto_localization_1984}: for broadband stimuli near the median plane ($|\alpha| \leq 30^\circ$), lateral responses are dominated by interaural cues, and show limited inter-individual variability~\cite{middlebrooks_sound_2015}. We therefore attribute individual lateral variability to motor uncertainty and fix the interaural noise parameters to literature values: $\sigma_{\text{itd}} = 0.569$ (dimensionless, corresponding to approximately one just-noticeable difference in ITD following the transformation of~\cite{reijniers_ideal-observer_2014}) and $\sigma_{\text{ild}} = 1.0$\,dB, consistent with ILD discrimination thresholds for broadband noise~\cite{yost_discrimination_1988}. Although individual variability in these parameters cannot be excluded, its impact is assessed empirically in Sec.~\ref{ssec:ident}. 

Having fixed $\sigma_{\text{itd}}$ and $\sigma_{\text{ild}}$, the free parameter vector reduces to $\boldsymbol{\theta} = \{\kappa_m, \sigma_{\text{mon}}, \sigma_{\text{prior}}\}$, which capture the primary sources of individual variability in static listening conditions. This reduction, combined with the assumption of isotropic motor noise (see Eq.~\ref{eq:response}), enables a sequential two-stage fitting procedure: a first stage restricted to lateral responses to estimate $\kappa_m$, and a second stage over all directional responses to estimate $\sigma_{\text{mon}}$ and $\sigma_{\text{prior}}$. The lateral likelihood is therefore defined over the horizontal angle $\alpha$ only:
\begin{equation}
    \mathcal{L}_\mathrm{lat}(\kappa_m) = \prod_{n=1}^{N} p(\hat{\alpha}^{*}_n \mid \alpha_n, \kappa_m),
    \label{eq:lateral_likelihood}
\end{equation}
where the observed responses $\hat{\dir}^*$ and target directions $\dir$ are omitted for readability. Because the interaural parameters are fixed, $\kappa_m$ is the sole free parameter and captures the residual lateral response variance not explained by the sensory model. In the first stage, $\hat{\kappa}_m$ is estimated by maximising $\mathcal{L}_\mathrm{lat}$ through optimisation.

In the second stage, $\hat{\kappa}_m$ is held fixed, and the remaining parameters $\sigma_\mathrm{mon}$ and $\sigma_\mathrm{prior}$ are jointly estimated by maximising the full-sphere likelihood over the complete set of directional responses:
\begin{align}
    \mathcal{L}_\mathrm{full}(\sigma_\mathrm{mon}, \sigma_\mathrm{prior} \mid \hat{\kappa}_m) 
    =& \nonumber\\
    \prod_{n=1}^{N} p(\hat{\dir}^{*}_{n} \mid \dir_n, \sigma_\mathrm{mon}, 
    \sigma_\mathrm{prior}, \hat{\kappa}_m).
    \label{eq:full_likelihood}
\end{align}
Note that $\hat{\kappa}_m$ is estimated under a von~Mises model on the circular lateral angle in the first stage, and subsequently used as the concentration parameter of the von~Mises--Fisher distribution on the unit sphere in the second stage: this is consistent since the von~Mises distribution is the restriction of the von~Mises--Fisher distribution to the circle.

\subsubsection{Likelihood optimisation and model comparison}
\label{sssec:optimisation}

For likelihood optimisation (Eq.~\ref{eq:full_likelihood}), we employ Bayesian Adaptive Direct Search~\cite[pyBADS,][]{acerbi_practical_2017}, a derivative-free method suited to objectives whose evaluations are noisy due to Monte Carlo approximation (see~\cite{barumerli_frambi_2025} for implementation details). The resulting fitted likelihoods enable hypothesis testing by quantifying how well model variants explain the observed data: variants encoding different mechanistic assumptions can be directly compared on this common scale~\cite{wilson_ten_2019}. However, more complex models will always explain the data better, and participants may contribute different numbers of trials, so a raw likelihood comparison would favour complexity and be sensitive to dataset size. To correct for both, we used the Bayesian Information Criterion~\cite[BIC;][]{vrieze_model_2012}, which penalises each additional free parameter by the log of the number of trials:
\begin{equation}
    \mathrm{BIC} = k \ln(N) - 2 \ln \mathcal{L}(\hat{\boldsymbol{\theta}}),
    \label{eq:bic}
\end{equation}
where $k$ is the number of free parameters, $N$ is the number of trials, and $\mathcal{L}_{\textrm{full}}(\hat{\boldsymbol{\theta}})$ is the likelihood at the estimated parameters. BIC differences across variants are interpretable on an absolute scale~\cite{kass_bayes_1995}, providing a principled criterion for model selection that accounts for both fit quality and model complexity. Note that, if $N$ and $k$ are fixed across all participants and model variants, the likelihood yields equivalent model rankings to BIC.

\subsection{Likelihood stability and parameter identification}
\label{ssec:param_id}

We first validate the statistical fitting procedure described above using simulated localisation responses. To this end, we simulate localisation responses as described in Sec.~\ref{sec:model} for 28 sets of uncorrelated ground-truth parameters spanning the expected range of individual variability, and investigate how well the parameters can be recovered by the fitting procedure. Parameter values are sampled to return comparable localisation performance (see Sec.~\ref{ssec:ident}).
Localisation responses are simulated using the KEMAR HRTF set from the SONICOM dataset~\cite{engel_sonicom_2023}, presenting 33 target directions with 3 repetitions each (99 trials), matching the minimum trial count in the experimental data (see Sec.~\ref{sssec:behav}). These configurations are reused across both the likelihood stability analysis and the parameter recovery analysis described below. 

First, we determine the number of Monte Carlo samples required for stable parameter estimation. We conduct a stability analysis by computing negative log-likelihood (NLL) variability across the 28 parameter configurations. We systematically vary the number of repetitions (3, 6, 9) for the 33 target directions (i.e. 99, 198 or 297 single trials) and the number of Monte Carlo samples for likelihood approximation (50--500). Then, we validate the fitting procedure by conducting a parameter recovery analysis using the simulated localisation responses. Recovery quality is quantified using the Pearson correlation coefficient $r$ and mean bias (defined as the average signed difference between recovered and ground-truth values; positive values indicate systematic overestimation).

\subsection{Behavioural Data and localisation metrics}
\label{sssec:behav}
We evaluate the model on 33 participants (aged 18- 43, median 26) who provided written informed consent under ethics approval (Imperial College London SETREC reference: 7046527, Declaration of Helsinki). All had their HRTFs measured as part of the SONICOM dataset~\cite{engel_sonicom_2023} prior to testing.

The data were collated from three experiments conducted in the same sound-insulated room, using broadband flat-spectrum sounds presented in static conditions, with an identical protocol (stimuli, apparatus, and familiarisation procedure as described in~\cite{daugintis_perceptual_2026}). Each experiment included a common baseline condition: static localisation of broadband sounds rendered binaurally with individual, acoustically measured HRTFs. Selecting only these trials yielded a variable number per participant: 9 participants contributed 297 trials (33 directions $\times$ 9 repetitions), 6 contributed 198 trials (6 repetitions), and the remainder contributed 99 trials (3 repetitions), all azimuths, with elevations down to $-30^\circ$ elevation.

Localisation performance is assessed using three standard metrics in the horizontal-polar coordinate system~\cite{middlebrooks_virtual_1999}. Lateral root mean squared (RMS) error (LE) quantifies left-right accuracy across all source directions. Local polar RMS error (PE) quantifies up-down and front-back accuracy for near-median-plane sources (lateral angles within $\pm30^\circ$), restricted to responses within the correct front-back hemisphere (polar errors below $90^\circ$) to exclude gross confusions. Quadrant error rate (QE) measures the rate of such gross front-back confusions, defined as the proportion of responses falling in the wrong hemisphere (polar errors above $90^\circ$) for the same near-median-plane subset as in PE.

\subsection{Full-sphere template interpolation}
\label{ssec:interpolation}

\begin{figure}
    \centering
    \includegraphics[width=1.0\linewidth]{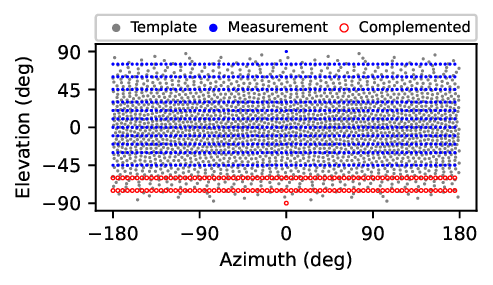}
    \caption{HRTF measurement and interpolated grid visualised over spherical coordinates. Blue points show the original measured directions of the G.R.A.S. KEMAR HRTF dataset~\cite{engel_sonicom_2023}; red circles indicate complemented directions in the bottom hemisphere where cues are extrapolated in the first step of the two-step interpolation method by mirroring measured directions across the horizontal plane (see Sec.~\ref{ssec:interpolation}). Grey points show the full template grid onto which all features are subsequently interpolated.}
    \label{fig:grid}
\end{figure}

Because the template represents the listener's learned internal representation of the auditory space, it is reasonable to assume that it is spatially continuous and full-spherical. The HRTFs from which the template is computed are spatially discrete in almost all cases, and often lack data around the South Pole (i.e. bottom directions) due to mechanical restrictions common to measurement systems (see Fig.~\ref{fig:grid} for an example based on an HRTF dataset from the SONICOM database~\cite{engel_sonicom_2023}). The template computation, hence, requires interpolation and extrapolation, we first compute the template features for the source positions for which HRTFs are available and then interpolate to a quasi-uniform spherical grid with $T = 2{,}112$ points and an average angular spacing of $4^\circ$ (a spherical $n$-design of degree~64~\cite{Graf2013, Graf2026}).

\begin{figure}
    \centering
    \includegraphics[width=1\linewidth]{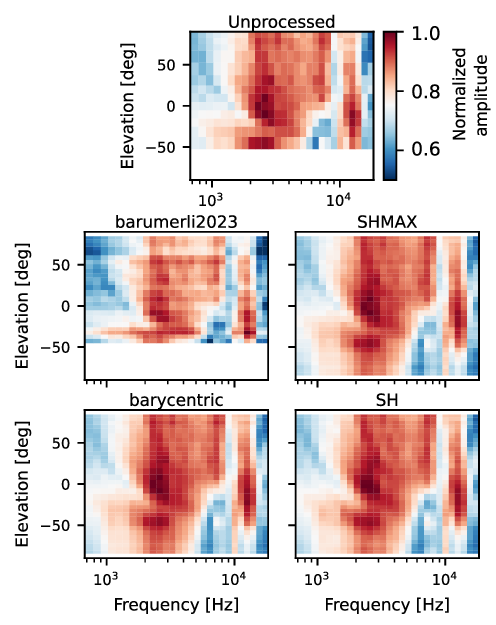}
    \caption{Patterns of spectral amplitudes cues on the median plane (left ear) originating from unprocessed G.R.A.S KEMAR HRTFs~\cite{engel_sonicom_2023} (top) and four interpolation methods. Each subplot is independently normalised to the maximum value to highlight spatial structure.}
    \label{fig:shinterp}
\end{figure}

An initial validation of the fitting procedure revealed systematic convergence failures attributable to the template interpolation of the original model (see Suppl.~\ref{supsec:temp_dist}). Specifically, the original interpolation method, termed here~\texttt{barumerli2023}, retains template features only above the minimum elevation $\epsilon_\mathrm{min}$ available in the measurement grid, so the model assigns no probability mass to a substantial portion of the sphere. Moreover, it introduces large interpolation errors close to the minimum elevation angle (see Fig.\ref{fig:shinterp}), biasing the posterior distribution even in the supported region.
We exploit this structural limitation as an opportunity to illustrate how model-based statistical analysis enables hypothesis testing: by comparing the original approach against three alternative interpolation methods providing full-sphere coverage, the BIC measure (see Eq.~\ref{eq:bic}) determines which best accounts for the observed localisation behaviour.
Importantly, the comparison is conducted on the same dense SONICOM measurement grid for all four methods, so that observed differences in localisation behaviour are attributable solely to the mathematical representation of the templates rather than to measurement density. Using a subset would introduce an additional variable that would confound this comparison.
\\ \\
\textit{Regularised Spherical Harmonic Interpolation}\label{sec:sh_interpolation}
\vspace{2mm}
\\
This method applies a regularised spherical harmonic (SH) transform of order $N_{SH}=15$~\cite[Section~3.6]{Rafaely2019} to compute the $(N_{SH}+1)^2$ SH coefficients $\boldsymbol{x}_{nm} = [x_{0,0}, x_{1,-1}, x_{1,0}, x_{1,1}, \hdots, x_{N_{SH},N_{SH}}]$ of degree $n$ and order $m$
\begin{align}
    \boldsymbol{x}_{nm} = \boldsymbol{Y}^\dagger_S\boldsymbol{x}_S\,
    \label{eq:SHT}
\end{align}
where $\boldsymbol{x_S} = [x_{\dir_1}, x_{\dir_1}, \hdots, x_{\dir_S},]^\mathrm{T}$ denotes any of the features introduced in Eq.~(\ref{eq:features}) at the $S$ source positions for which the HRTF is available. $\boldsymbol{Y}_S \in \mathbb{R}^{S \times (N_{SH}+1)^2}$ is the matrix containing the real-valued SH basis functions~\cite{Zotter2019} for all source positions, orders and degrees. $(\cdot)^\dagger$ denotes the regulated pseudo-inverse using $\mathrm{diag}(\boldsymbol{\lambda}_n) \in \mathbb{R}^{(N_{SH}+1)^2 \times (N_{SH}+1)^2}$ for regularisation, with $\boldsymbol{\lambda}_n$ being 0 (no regularisation) for orders $n\leq2$ and 4 (regularisation) otherwise. The interpolation to the template grid is then realised by
\begin{align}
    \boldsymbol{x}_T = \boldsymbol{Y}_T\boldsymbol{x}_{nm}
    \label{eq:ISHT}
\end{align}
with $\boldsymbol{Y}_T \in \mathbb{R}^{T \times (N_{SH}+1)^2}$ holding the basis functions for the positions of the template grid.

The original implementation \texttt{barumerli2023} kept only feature values for elevation angles $\epsilon > \epsilon_\mathrm{min}$. This is required because $\boldsymbol{x}_T$ exhibits large extrapolation errors outside this region (cf.~\cite{Ahrens2012a} - see Supplementary materials~\ref{supsec:temp_dist}). It is the default method used in the original MATLAB version of the model and was kept for backwards compatibility.
\\
\\
\textit{Two-step interpolation}
\vspace{2mm}
\\
To extrapolate missing data at low elevation, we use the low-order SH interpolation suggested by Ahrens et~al.~\cite{Ahrens2012a} based on Eqn. (\ref{eq:SHT}-\ref{eq:ISHT}) with the following two differences. First, the order of the SH transform in (\ref{eq:SHT}) is determined by finding the maximum order for which $\mathrm{cond}(\boldsymbol{Y}^\dagger)<12.25$ is satisfied using $\lambda_n=n \cdot 10^{-2}$ for regularisation. The condition number $\mathrm{cond}(\cdot)$ is given by the ratio of the largest and smallest eigenvalues and provides a quantity on the stability of the transformation: given a fixed input, the higher the condition number, the higher will be the output variability. The selected limits for the condition number and regularisation are appropriate for irregular sampling grids tested in Bau et~al.~\cite{bau_estimation_2022}.

Second, the interpolation described in (\ref{eq:ISHT}) is evaluated for the source positions $S^\prime$ to complement missing data in the sampling grid. These source positions are defined as suggested by Ahrens et al.~\cite{Ahrens2012a}: the minimum elevation $\epsilon_\mathrm{min}$ contained in $S$ is found, which is often between $-30^\circ$ and $-60^\circ$, and all source positions with an elevation larger than $-\epsilon_\mathrm{min}$ are mirrored at the horizontal plane to complement the sampling grid.

In the last step, the features that are now available on the complemented sampling grid $S \cup S^\prime$ are interpolated to the template grid using either SH interpolation or Barycentric interpolation, which uses data from up to three surrounding sampling points selected based on a triangulation of the sampling grid~\cite{Gamper2013}. In this case, we refer to this method as \texttt{barycentric}. If using SH interpolation, the maximum order can be determined as in the first step (termed \texttt{SH}) or be freely specified while still using the regularisation described above (termed \texttt{SHMAX}). For the SONICOM grid used here, this yields a maximum order of 11 for SH; SHMAX uses a fixed order of 44.

Compared to the default, the two-step approach offers a full-spherical template. However, the spatial resolution of the interpolated data below $\epsilon_\mathrm{min}$ might be limited and depend on the SH order used for extrapolation in the first step.

\subsection{Python implementation}
\label{sec:implementation}
The model is distributed as \texttt{bayesian\_listener}, an open-source Python package released under the EUPL 1.2 license installable via \texttt{pip install bayesian\_listener}. The documentation is linked from the source code available at \url{https://github.com/robaru/bayesian\_listener}.
Core dependencies include \texttt{numpy} and \texttt{scipy} for numerical operations, \texttt{pyfar} and \texttt{spharpy} for spherical harmonics interpolation and coordinate handling, \texttt{sofar} for reading SOFA-formatted HRTF datasets, and \texttt{pybads}~\cite{acerbi_practical_2017} for derivative-free parameter optimisation.

The package exposes a single high-level class that encapsulates the three model stages (feature extraction, Bayesian inference, and response noise), accepting either a SOFA file or a precomputed template array. All free parameters can be set independently, or maximum likelihood estimation is available directly through the class interface via the two-stage procedure described in Sec.~\ref{sec:fitting}.  Application Programming Interface (API) documentation is generated with Sphinx, and unit tests are implemented via \texttt{pytest} under Python~3.10+ on Linux, macOS, and 
MS Windows.

\section{Results}

Results are organised following the Bayesian workflow of~\cite{wilson_ten_2019}: we first validated the Python implementation against the original MATLAB model, then assessed template reconstruction quality, characterised the fitting procedure through likelihood stability and parameter recovery on simulated localisation responses, and finally applied it to behavioural data from 33 participants, evaluating fit quality through posterior predictive checks and interpolation comparison based on BICs.

\begin{figure}[!t]
    \centering
    \includegraphics[width=0.8\linewidth]{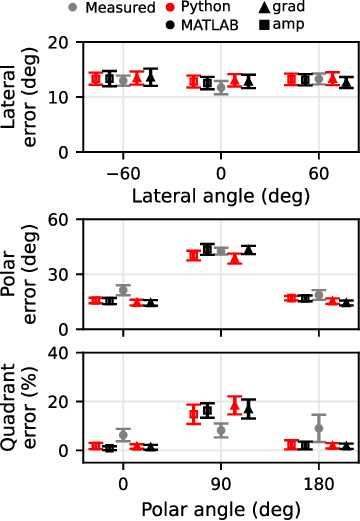}
    \caption{Replication of Figure 5 from~\cite{barumerli_bayesian_2023}, validating the Python implementation (red) against the original MATLAB model (black), comparing sound-localisation performance against behavioural data (grey) from~\cite{majdak_3-d_2010}. Model parameters are taken directly from the original study; error bars denote standard errors across the five listeners therein. Squares and triangles denote the two alternatives for characterising monaural features proposed in the original study: spectral amplitude and positive spectral gradient monaural cues, respectively~\cite{barumerli_bayesian_2023}.}
    \label{fig:repro}
\end{figure}

\subsection{Evaluation of model implementation}

We first verified that the Python implementation reproduces the behaviour of the original MATLAB model before proceeding to the likelihood-based analysis. Figure~\ref{fig:repro} replicates Figure~5 of \cite{barumerli_bayesian_2023} using the five subjects from~\cite{majdak_3-d_2010} and individual HRTFs, with model parameters obtained via the initial iterative procedure of the original study. The agreement between the two implementations is close across all three performance metrics and across listeners. A discrepancy was observed at $90^\circ$ polar angle, where the Python implementation predicted a slightly smaller polar error than MATLAB for both the amplitude ($40.2 \pm 2.6^\circ$ vs.\ $43.5 \pm 3.0^\circ$) and gradient ($38.5 \pm 2.7^\circ$ vs.\ $43.2 \pm 2.3^\circ$) variants (mean $\pm$ standard error over five participants). This is attributable to the denser template grid in the Python implementation, arising from a different quasi-uniform grid generation algorithm producing a slightly finer spatial sampling at the same target angular spacing of $4^\circ$. Given the overlapping standard errors across five listeners and the consistency across all metrics and directions, this difference does not indicate a systematic implementation discrepancy, though we note that a formal equivalence test is underpowered at this sample size. This informal equivalence check establishes the Python package as a valid basis for the likelihood-based analyses that follow.

Since the two spectral feature variants yield closely comparable predictions across all metrics for flat and broadband sounds~\cite{barumerli_bayesian_2023}, the remainder of this study focuses on spectral amplitude features; the gradient variant remains available in the Python implementation.

\subsection{Template reconstruction quality}

Prior to fitting and model analysis, we examined how the choice of template interpolation method affects the spectral features on which localisation predictions depend. Figure~\ref{fig:shinterp} illustrates the spectral magnitude patterns on the median plane for the G.R.A.S.\ KEMAR manikin HRTF across the four interpolation methods, with \texttt{barumerli2023} showing visible distortions relative to the unprocessed HRTF. Such a result is consistent with its large root mean squared error (RMSE) computed across source positions, frequencies and 33 participants (Sec.~\ref{sssec:behav}), which averaged $2.03\pm0.04$ dB (mean $\pm$ standard error): up to roughly an order of magnitude larger than the best full-sphere methods. Among the latter, \texttt{barycentric} and \texttt{SHMAX} achieved the lowest errors ($0.24\pm0.00$ and $0.23\pm0.01$ dB), while \texttt{SH} interpolation yielded a moderately higher RMSE of $0.49\pm0.02$ dB, reflecting frequency-dependent smoothing from the condition-number constraint on SH order~\cite{ben-hur_binaural_2021}. Overall, full-sphere spatial coverage emerges as a more important determinant of reconstruction fidelity than the choice of interpolation algorithm, at least for the dense SONICOM measurement grid used here.

\subsection{Likelihood stability and parameters identification}
\label{ssec:ident}
We first determined the number of Monte Carlo samples required for stable likelihood approximation, then assessed whether the free parameters can be reliably recovered from simulated localisation responses.

For such simulations, ground-truth parameters spanned the expected range of individual variability ($\sigma_{\text{m}} \in [5, 16]^\circ$, $\sigma_{\text{mon}} \in [2, 15]$ dB, $\sigma_{\text{prior}} \in [5, 90]^\circ$), producing simulated localisation performance covering the range observed in static localisation scenarios (LE: $10.9^\circ$, $[5.4, 16.8]^\circ$; PE: $31.1^\circ$, $[10.8, 43.0]^\circ$; QE: $12.4$\%, $[0.0, 27.2]$\%; mean and $[\min, \max]$ across 28 configurations)~\cite{carlile_nature_1997, middlebrooks_virtual_1999, majdak_3-d_2010}. 

Likelihood stability was primarily governed by the number of Monte Carlo samples rather than the number of trial repetitions, indicating that each trial contributes information with equal efficiency regardless of how many times each direction is presented. Increasing from 50 to 200 samples reduced per-trial likelihood standard deviation by approximately $53\%$ (from $4.31\pm3.86$ to $2.01\pm0.85$, mean~$\pm$~standard error across the 28 configurations), with diminishing returns beyond this point (see Supplementary Material~\ref{supsec:lik_stab}). We therefore selected 200 Monte Carlo samples as a sufficient approximation for all subsequent analyses. 

Parameter identification was assessed by comparing ground-truth and recovered values through likelihood optimisation using Pearson correlation coefficients ($r$) and mean bias (see Supplementary Material~\ref{supsec:par_rec} for visualisation of single estimates). The motor noise parameter $\sigma_{m}$ showed agreement between ground-truth and recovered values ($r = 0.97$, $ p < 0.001$, mean bias $= -0.1^\circ$), confirming that the two-stage profile likelihood decomposition effectively isolates the sensorimotor component from the spectral processing parameters. The spectral noise $\sigma_\mathrm{mon}$ and prior width $\sigma_\mathrm{prior}$ exhibited strong recovery ($r = 0.85$, $ p < 0.001$ and $r = 0.84$, $ p < 0.001$, respectively), with positive biases (mean bias: $+3.0$ and $+19.9^\circ$). For $\sigma_\mathrm{prior}$, the positive bias occurs at large values where the prior approaches a uniform distribution over elevation, making large parameter values behaviourally indistinguishable. For $\sigma_\mathrm{mon}$, recovery degrades at large values corresponding to highly imprecise localisers, whose near-chance polar responses carry insufficient information to constrain the estimate. Crucially, recovered parameters were uncorrelated ($r = -0.01$, $p = .95$ for $\sigma_\mathrm{m}$ vs.\ $\sigma_\mathrm{mon}$; $r = -0.16$, $p = .42$ for $\sigma_\mathrm{mon}$ vs.\ $\sigma_\mathrm{prior}$), confirming that the optimisation procedure introduces no spurious trade-offs between parameter estimates.

To assess whether parameter recovery preserved behaviourally relevant information, we compared standard localisation metrics computed from the ground-truth and recovered parameter sets. Lateral errors showed a strong positive correlation between ground-truth and recovered predictions ($r = 0.96$, $p < 0.001$, mean difference $\pm$ standard deviation $-0.04^\circ \pm 0.79^\circ$), while polar errors and quadrant error rates showed a smaller, albeit still strong, correlation ($r = 0.72$, $p < 0.001$ and $r = 0.80$, $p < 0.001$ respectively), with small mean differences ($-0.34^\circ \pm 6.92^\circ$ and $2.08\% \pm 6.75\%$). These results indicate that although individual parameter estimates carry uncertainty, particularly for $\sigma_{\text{prior}}$ and $\sigma_{\text{mon}}$ at large values, this uncertainty does not propagate into behaviourally meaningful errors: localisation metrics computed from recovered parameters closely match those from ground-truth parameters, confirming that the fitted models capture the relevant structure of localisation performance.

\subsection{Parameter estimates from behavioural data}
\begin{figure}[t]
    \centering
    \includegraphics[width=1\linewidth]{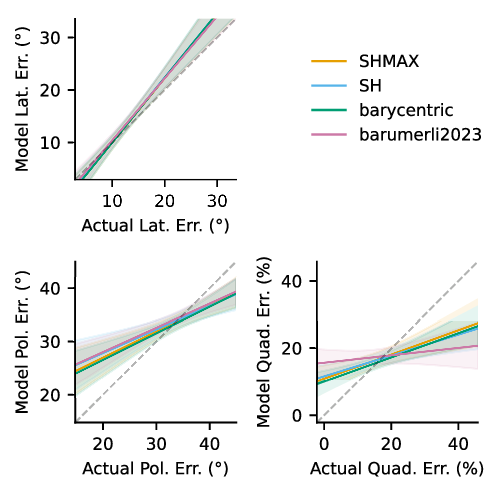}
    \caption{Model predictions of localisation metrics across the range of observed behaviour, serving as posterior predictive checks for the four interpolation methods. Each panel shows a different localisation metric (lateral error, polar error, and quadrant error rate), with actual behavioural responses on the $x$-axis and simulated localisation responses generated from fitted model parameters on the $y$-axis. The dashed identity line indicates perfect prediction; points above it indicate model overestimation. Linear regressions with standard-error ribbons summarise the trend for each interpolation method. The key result is in the quadrant error rate panel: \texttt{barumerli2023} fails to track individual differences ($r = 0.19$, $p = .30$), whereas full-sphere methods succeed ($r = 0.51$--$0.60$, $p<.05$), consistent with the template artefacts at low elevations introduced by the original interpolation method.}
    \label{fig:posteriorchecks}
\end{figure}

Table~\ref{tab:metrics} reports group-level parameter estimates and fit statistics for all participants and interpolation methods. The motor noise parameter $\sigma_\mathrm{m}$, estimated independently on the lateral localisation data, ranged from $2.2^\circ$ to $35.7^\circ$ across participants (median $= 9.7^\circ$). Within-participant estimates were near-identical across methods (Table~\ref{tab:metrics} - top panel), confirming that this parameter captures a listener-specific sensorimotor component independent of the template representation. The range of fitted values is consistent with the sensorimotor scatter reported previously \cite{barumerli_bayesian_2023, daugintis_classifying_2023}, with the wider spread here reflecting the larger participant pool ($N = 33$ vs.\ $N = 5$).

To verify that fixing $\sigma_\mathrm{itd}$ and $\sigma_\mathrm{ild}$ did not absorb individual differences in binaural sensitivity into $\hat{\sigma}_m$, we repeated the Stage~1 estimation for all 33 participants across the full $3\times 3$ grid. Across the eight non-default conditions, the correlation between perturbed and default $\hat{\sigma}_m$ estimates exceeds $r = 0.998$ in every case, with a mean absolute deviation of $0.30^\circ$ and a maximum of $2.05^\circ$, both small relative to the inter-individual range of $2.2^\circ$--$35.7^\circ$. To further quantify the contribution of $\sigma_\mathrm{itd}/\sigma_\mathrm{ild}$ variation to lateral accuracy, we evaluated the ITD$+$ILD-only model ($\sigma_m = \infty$) across the same grid. The resulting $LE$ values have a mean of $2.7^\circ \pm 1.4^\circ$, and the spread attributable to the full grid does not exceed $1.94^\circ$ per participant (mean $0.51^\circ$), compared to an observed inter-individual $LE$ range of $3.0^\circ$--$34.7^\circ$. These results confirmed that individual differences in $\sigma_\mathrm{itd}$ and $\sigma_\mathrm{ild}$ within physiologically plausible ranges cannot account for the observed spread in lateral pointing accuracy, and that the estimation procedure correctly attributes this variance to $\kappa_m$.

The jointly optimised $\sigma_\mathrm{mon}$ and $\sigma_\mathrm{prior}$ exhibited substantial individual variability (Table~\ref{tab:metrics}), consistent with known listener-specific differences in spectral processing and spatial expectations. To compare fitted parameter values across interpolation methods, we used Spearman rank correlations ($\rho$) to avoid assumptions on the marginal distributions. Importantly, the two parameters responded differently to the choice of interpolation method. Spectral noise did not differ significantly across methods (Friedman $\chi^2(3) = 5.33$, $p = .149$), though participant-level estimates were more strongly correlated between high-resolution methods (\texttt{SHMAX}--\texttt{barycentric}: $\rho = 0.81$, $p < .001$) and weakest against the original approach (\texttt{SHMAX}--\texttt{barumerli2023}: $\rho = 0.45$, $p = .008$), suggesting that higher-fidelity templates yield more consistent sensory parameter estimates. Prior width, by contrast, showed a significant method effect (Friedman $\chi^2(3) = 10.91$, $p = .012$), with considerable within-participant variability across methods (median range $= 22.9^\circ$). This pattern reveals a confound between template quality and prior estimation: when template quality is distorted, as in \texttt{barumerli2023}, the optimiser tightens $\sigma_{\text{prior}}$ to compensate for the degraded sensory evidence, meaning that $\sigma_{\text{prior}}$ cannot be interpreted as a pure listener property in the presence of template artefacts. This further motivates the use of full-sphere interpolation methods, which avoid this confound by providing undistorted templates.

Our $\sigma_\mathrm{mon}$ and $\sigma_\mathrm{prior}$ estimates are substantially larger than those reported in the original studies \cite{barumerli_bayesian_2023, daugintis_classifying_2023}. First, we fitted $\sigma_\mathrm{prior}$, resulting in a broader mean, whereas in the original study it was fixed at $11.5^\circ$. Our result is consistent with~\cite{ege_perceived_2019}, where they showed that listeners adapt the width of their spatial prior to the statistical distribution of source positions: our stimulus set spanned a larger elevation range, which would be expected to yield a wider prior estimate. Second, the discrepancy on $\sigma_\mathrm{mon}$ is expected: original studies employed an ad hoc estimation method on a small set of perceptually trained participants~\cite{majdak_3-d_2010}, whereas the present work uses maximum likelihood estimation over a larger cohort whose members completed only a familiarisation session~\cite{daugintis_perceptual_2026}. Despite these parameter differences, the likelihood optimisation yields statistically interpretable estimates for parameters previously fixed in the ad hoc procedure (e.g. $\sigma_\mathrm{prior}$), and their consistency with observed localisation metrics (Sec.~\ref{sssec:postchecks}) demonstrates the greater inferential power of the maximum likelihood framework over heuristic alternatives.

\begin{table}[]
    \centering
\begin{tabular}{lccc}
\multicolumn{4}{c}{}\\
\multicolumn{4}{c}{Estimated parameters from behavioural data}\\
\hline

Method & $\sigma_\mathrm{m}$ & $\sigma_\mathrm{mon}$ & $\sigma_\mathrm{prior}$\\
\hline
\texttt{barumerli2023} & $12.5\pm1.4$ & $9.5\pm0.7$ & $55.0\pm6.2$ \\
\texttt{SHMAX} & $12.5\pm1.5$ & $10.4\pm0.9$ & $69.0\pm9.0$ \\
\texttt{barycentric} & $12.5\pm1.5$ & $9.5\pm0.6$ & $73.5\pm10.0$  \\
\texttt{SH} & $12.5 \pm 1.5$ & $10.1\pm0.8$ & $78.1\pm9.6$  \\
\hline
\end{tabular}
    
\begin{tabular}{lccccc}
\multicolumn{4}{c}{}\\
\multicolumn{4}{c}{Estimated behavioural metrics}\\
\hline
Method & LE ($^\circ$) & PE ($^\circ$) & QE (\%) \\
\hline
Actual & 12.7 $\pm$ 0.9 & 33.8 $\pm$ 1.1 & 15.1 $\pm$ 1.8 \\
\texttt{barumerli2023} & 13.6 $\pm$ 1.2 & 34.3 $\pm$ 0.7 & 17.4 $\pm$ 1.1 \\
\texttt{SHMAX} & 13.3 $\pm$ 1.2 & 33.7 $\pm$ 0.8 & 16.4 $\pm$ 1.3 \\
\texttt{barycentric} & 13.3 $\pm$ 1.2 & 33.4 $\pm$ 0.8 & 15.5 $\pm$ 1.1 \\
\texttt{SH} & 13.5 $\pm$ 1.2 & 33.9 $\pm$ 0.8 & 16.4 $\pm$ 1.1 \\
\hline

\multicolumn{4}{c}{}\\
\multicolumn{4}{c}{Correlations between model and participant metrics}\\
\hline
Method & LE ($^\circ$) & PE ($^\circ$) & QE ($\%$) \\
\hline
\texttt{barumerli2023} & 0.89*** & 0.65*** & 0.19 \\
\texttt{SHMAX} & 0.89*** & 0.63*** & 0.52** \\
\texttt{barycentric} & 0.89*** & 0.65*** & 0.60*** \\
\texttt{SH} & 0.89*** & 0.57*** & 0.51** \\
\hline
\end{tabular}
    \caption{Comparison of template interpolation methods across estimated parameters (top panel), and behavioural estimation accuracy (values reported as mean $\pm$ standard error over participants). The behavioural metrics (middle panel) compare actual localisation performance against simulated localisation responses from fitted models, including lateral RMS error (LE), polar RMS error (PE), and quadrant error (QE). Correlations (Pearson coefficient $r$, bottom panel) indicate each method's ability to predict individual differences in localisation performance across dimensions. Significance levels: *** $p < 0.001$, ** $p < 0.01$.}
    \label{tab:metrics}
\end{table}

\subsubsection{Posterior predictive checks}
\label{sssec:postchecks}
Posterior predictive checks assess model fit by generating localisation responses from the fitted model and comparing them to the observed experimental data; systematic discrepancies indicate where the model fails to capture the structure of localisation behaviour~\cite{wilson_ten_2019}. To this end, for each participant and interpolation method, the model generated responses to the same target directions presented in the experiment, averaged over 200 Monte Carlo repetitions. The obtained localisation metrics were compared against those obtained from the actual experimental data (Table~\ref{tab:metrics}, Fig.~\ref{fig:posteriorchecks}). We evaluated lateral RMS error ($\mathrm{LE}$), local polar error ($\mathrm{PE}$), and quadrant error rate ($\mathrm{QE}$), following the definitions of \cite{middlebrooks_virtual_1999}.

Across all methods, the model reproduced group-level behavioural metrics with no significant bias (Wilcoxon $p > .27$ for $\mathrm{LE}$, $\mathrm{PE}$, and $\mathrm{QE}$; Table~\ref{tab:metrics}). Individual differences in lateral error were captured with uniformly high fidelity ($r = 0.89$, all methods), confirming that the two-stage fitting procedure preserves listener-specific sensorimotor variability. Polar error correlations were moderate-to-strong across methods ($r = 0.57$--$0.65$), while quadrant error rates revealed the clearest differentiation: \texttt{barycentric} and \texttt{SHMAX} interpolation tracked individual QE differences ($r = 0.60$ and $r = 0.52$, respectively), whereas \texttt{barumerli2023} did not ($r = 0.19$, $p = .30$). This failure is consistent with the template artefacts at low elevations introduced by the original method, which distort the posterior distribution in regions where front--back confusions are resolved.

We also examined two accuracy metrics that fell outside the model's scope. The lateral accuracy bias showed a systematic offset of approximately $3^\circ$ (Wilcoxon $p = .002$), which the model cannot reproduce by construction since the response noise is assumed zero-mean and isotropic (Eq.~\ref{eq:response}); this is consistent with a minor violation of the isotropy assumption not captured by the current model formulation. The polar accuracy showed no significant differences between model predictions and data ($p > .26$), but the model did not capture across-participant variability in this metric ($r < 0.22$, $p > .21$), indicating a structural limitation of the current formulation in accounting for individual vertical biases.

\subsection{Model comparison}
\label{ssec:comparison}
\begin{figure}
    \centering
    \includegraphics[width=1\linewidth]{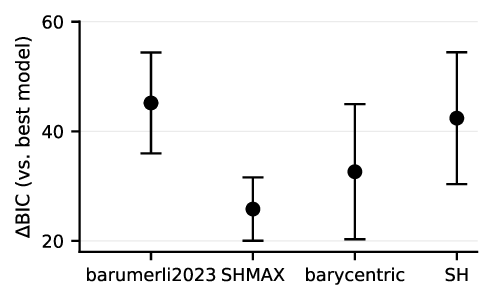}
    \caption{Model fit quality across template interpolation methods. Points and error bars show mean $\Delta\mathrm{BIC}$ ($\pm$ standard error over 33 participants) relative to the best-fitting method for each participant. Lower values indicate better fits.}
    \label{fig:bic_comparison}
\end{figure}

\begin{table}
    \centering
    \begin{tabular}{lcccc}
    \hline Method & $\Delta$BIC & Wins & Very strong \\ \hline
    SHMAX & -19.4 $\pm$ 10.1 & 19/33 & 17  \\ 
    barycentric & -12.5 $\pm$ 16.7 & 18/33 & 15  \\
    SH & -2.8 $\pm$ 15.3 & 15/33 &  13  \\ 
    \hline
    \end{tabular}
\caption{Pairwise model comparison against the \texttt{barumerli2023} interpolation method. $\Delta$BIC: mean $\pm$ standard error over all participants of the BIC difference relative to \texttt{barumerli2023} (negative values favour the alternative method). Wins: number of participants (out of 33) for whom the method achieved a lower BIC. Very strong indicates the number of participants with a large preference for the alternative method ($|\Delta\mathrm{BIC}| > 10$). }
\label{tab:bic_comparison}
\end{table}

We compared the four interpolation methods by computing $\Delta\mathrm{BIC}$ for each participant, defined as the difference between each method's BIC and the best-fitting method, with likelihoods evaluated using 500 Monte Carlo samples per trial to reduce estimation uncertainty (see Fig.~\ref{fig:bic_comparison}). Because raw BIC values vary widely across participants due to differences in overall localisation ability, $\Delta\mathrm{BIC}$ isolates the relative advantage of each method within each listener. Lower $\Delta\mathrm{BIC}$ values indicate better relative fit, with differences exceeding 10 conventionally regarded as strong evidence in favour of the better-fitting model~\cite{kass_bayes_1995}.

A Friedman test on $\Delta\mathrm{BIC}$ across methods did not reach significance ($\chi^2(3) = 5.04$, $p = .169$), indicating that no single method uniformly dominated across participants. Descriptively, \texttt{SHMAX} and \texttt{barycentric} achieved the lowest mean $\Delta\mathrm{BIC}$ ($25.8$ and $32.6$, respectively), while \texttt{SH} and \texttt{barumerli2023} showed larger deviations from the best-fitting method ($42.4$ and $45.2$) (see Fig.~\ref{fig:bic_comparison}). 

At the individual level, categorising $\Delta\mathrm{BIC}$ between \texttt{SHMAX} and \texttt{barumerli2023} according to \cite{kass_bayes_1995} yielded 17 participants with very strong evidence ($\Delta\mathrm{BIC} > 10$) and 2 with positive evidence ($2 < \Delta\mathrm{BIC} \leq 10$) favouring \texttt{SHMAX}, 4 with negligible differences ($|\Delta\mathrm{BIC}| \leq 2$), and 10 favouring \texttt{barumerli2023} (see Tab.~\ref{tab:bic_comparison}). A similar pattern held for \texttt{barycentric} versus \texttt{barumerli2023}, suggesting that once full-sphere coverage and adequate spectral bandwidth are ensured, the choice of interpolation algorithm yields no additional benefit for grids as dense as the SONICOM measurement set~\cite{engel_sonicom_2023}.

\section{Discussion}
This work presents two contributions, a methodological framework and an applied finding, with the aim of demonstrating the value of principled statistical methods in computational auditory modelling~\cite{pitt_when_2002, barumerli_frambi_2025}. Both rest on a Python re-implementation of the Barumerli et al.\cite{barumerli_bayesian_2023} model, whose close agreement with the original MATLAB version across all localisation metrics and listeners (Fig.\ref{fig:repro}) ensures that the differences reported reflect genuine perceptual model behaviour rather than implementation artefacts. The discussion examines each contribution in turn before illustrating how they converge.

A central requirement for any likelihood-based fitting procedure is that the free parameters can be reliably recovered from generated data~\cite{wilson_ten_2019}. 
Non-identifiability manifests in two distinct ways: a single parameter is poorly constrained by the data (low correlation between ground-truth and recovered values); multiple parameters trade off against each other, such that a change in one is compensated by another (high correlation between recovered parameters), yielding identical behavioural predictions despite different parameter combinations. Once detected, these limits can be managed by fixing non-identifiable parameters to independent literature values, or by incorporating additional behavioural data that selectively isolates their contributions.
In the lateral dimension, $\sigma_{\text{itd}}$, $\sigma_{\text{ild}}$, and $\kappa_m$ are jointly non-identifiable from directional responses alone. Fixing $\sigma_{\text{itd}}$ and $\sigma_{\text{ild}}$ to literature values resolved this, yielding strong recovery of $\kappa_m$ ($r = 0.97$). In the vertical dimension, $\sigma_{\text{mon}}$ and $\sigma_{\text{prior}}$ did not exhibit spurious correlation in the recovery analysis (Sec .~\ref {ssec:ident}), excluding potential trade-offs. However, their moderate recovery ($r = 0.85$ and $0.84$) reflects a fundamental model property: both parameters govern the vertical response distribution and become behaviourally indistinguishable when vertical localisation performance is poor (see supplementary material \ref{supsec:par_rec} for a visualisation), as observed in listeners without prior familiarisation with the experimental procedure~\cite{majdak_3-d_2010}. In such cases, the likelihood surface becomes flat with respect to these parameters, and the optimiser tends to overestimate both. Practitioners should therefore anticipate reduced reliability when fitting to listeners with degraded vertical localisation performance.

Model comparison via information criteria is a standard tool for adjudicating between competing hypotheses, yet it identifies the best-fitting model within the comparison set without guaranteeing absolute model adequacy, requiring multiple sources of evidence to evaluate where and why a model fails~\cite{wilson_ten_2019}. This limitation manifested concretely in our data: for 10 of 33 participants, \texttt{barumerli2023} achieved the best BIC (i.e. lowest value) despite its known template artefacts, because the optimiser exploited a narrower prior width (mean $\sigma_{\text{prior}}=55.0^\circ$ vs.\ $69.0$--$78.1^\circ$ for full-sphere methods) to compensate for degraded sensory evidence rather than exposing it as a model failure. Evaluating template quality solely on BIC would therefore have been insufficient, and the conclusion required converging evidence from parameter recovery, posterior predictive checks, and acoustic reconstruction quality.

The original model formulation relied on a truncated template that discarded directions below the lowest measured HRTF elevation, avoiding the spurious spectral distortions that arise when spherical harmonic interpolation is evaluated at directions far outside the measured region. Such an assumption was never formally tested against alternatives~\cite{barumerli_bayesian_2023}. Applying the likelihood-based fitting to this interpolation scheme revealed systematic distortion in the bottom hemisphere attributable to the absence of acoustic measurement coverage in that area, a limitation invisible to the previous metric-matching procedure because the missing elevation range contributed little to aggregated scalar summaries. The likelihood function, by contrast, evaluates each individual trial, exposing regions where the posterior distribution becomes ill-defined~\cite{peel_fitting_2001}. This finding motivated the full-sphere interpolation comparison, which showed that spatial coverage and preservation of high-frequency spectral content are the primary determinants of template quality. Among full-sphere methods, \texttt{SH} interpolation at order 11 underperformed because the condition-number constraint acts as a spatial low-pass filter~\cite{ben-hur_spectral_2017}, attenuating spectral detail relevant to polar localisation, whereas the \texttt{SHMAX} approach avoided this by selecting the maximum order consistent with numerical stability for each grid separately. We therefore recommend treating the condition number and SH order as grid-specific diagnostics rather than applying a fixed order across datasets, as already noted by~\cite{bau_estimation_2022}. Once full-sphere coverage and adequate spectral bandwidth are ensured, the specific interpolation algorithm yields no additional benefit for grids as dense as the SONICOM measurement set.

Beyond evaluating interpolation methods, the posterior predictive checks (Sec.~\ref{sssec:postchecks}) revealed where the model succeeds and where it reaches its structural limits. The high fidelity of lateral error predictions across all methods was consistent with the dominance of interaural cues, a well-constrained dimension largely independent of template quality. The moderate agreement for polar errors and quadrant error rates reflected the identifiability limits of $\sigma_{\text{mon}}$ and $\sigma_{\text{prior}}$ discussed above. Two behavioural patterns fell outside the model's scope: the systematic lateral accuracy bias, which the model cannot reproduce since response noise is zero-mean by construction and likely reflects pointing apparatus calibration or motor asymmetries~\cite{daugintis_perceptual_2026}; and individual polar accuracy variability, suggesting that polar biases arise from factors such as asymmetric prior beliefs or idiosyncratic head posture not represented in the current formulation. This evidence delineates phenomena that future extensions could target.

Several additional limitations should be noted. First, the assumption of isotropic motor noise models response variance but not bias. Direction-dependent pointing biases are well documented for egocentric pointing methods comparable to the controller-based paradigm used here~\cite{bahu_comparison_2016, boyer_ear_2013}, and the small lateral bias observed in our data (Sec.~\ref{sssec:postchecks}) is consistent with these effects. However, parametrising a direction-dependent motor model would require a dedicated protocol isolating motor from perceptual error, which we leave for future work.
Second, the prior width $\sigma_{\text{prior}}$ showed limited identifiability and absorbed differences in template quality across interpolation methods, indicating that it did not purely reflect a stable listener trait; separating prior from likelihood contributions would require experimental designs with manipulated prior expectations or adaptive stimulus placement, which are beyond the current scope~\cite{ege_perceived_2019}. 
Third, the behavioural data were collated across three experiments with variable trial counts per participant (99--297 trials), potentially introducing heterogeneous measurement precision; while the minimum of 99 trials proved sufficient for parameter recovery, listeners with fewer repetitions carry larger estimation uncertainty. 
Finally, the large individual variability observed across all parameters and fit statistics underscores that sound localisation is shaped by idiosyncratic combinations of sensory precision, learned expectations, and sensorimotor factors~\cite{carlile_nature_1997, middlebrooks_virtual_1999, majdak_acoustic_2014}, and no single model formulation can be expected to capture all sources of variability.

While the newly proposed spatial cue interpolation improved the model fit, two aspects warrant further attention in future work. First, we did not formally evaluate the quality of the extrapolated spatial cues below $\epsilon_\mathrm{min}$. It has been shown that broadband onset times can be represented by very low SH orders \cite[Fig.~4]{Brinkmann2018c}, suggesting that this is also true for broadband ITDs $x_{itd}$ and ILDs $x_{ild}$. It is reasonable to assume that interpolating spectral cues $x_{mon}$ requires higher SH orders and that resulting interpolation errors might bias model predictions. Almost identically computed spectral cues were previously used to derive magnitude correction filters that improved the perceived source position and colouration in HRTFs interpolated with SH orders between 1 and 3~\cite{Arend2023b}. This indicates that spectral cues might be interpolatable with relatively low SH orders. Second, the interpolation approaches have so far been tested only for the SONICOM sampling grid, and the influence of the latter on interpolation error and model estimates could be the focus of future investigations on grids used in common HRTF databases.

A natural extension of the current analysis is full Bayesian posterior inference over model parameters via Markov chain Monte Carlo (MCMC) sampling rather than likelihood optimisation~\cite{bishop_pattern_2006}. Rather than returning point estimates, this approach yields a full posterior distribution over parameters, enabling probabilistic queries about latent variables and direct assessment of identifiability limits through the shape of the posterior. In addition, MCMC enables the possibility of hierarchical extensions for simultaneous inference over individual and group-level parameters~\cite{turner_bayesian_2013}, and joint fitting of heterogeneous data sources combining localisation responses with neurophysiological recordings~\cite{greif_role_2025} to constrain shared latent parameters where behavioural data alone are insufficient. However, the primary barrier is computational: the likelihood must be evaluated thousands of times during sampling, making MCMC currently prohibitive for datasets of this size, though recent advances in amortised inference may reduce this cost by training neural networks to approximate the posterior directly~\cite{cranmer_frontier_2020}.

The validated likelihood framework naturally extends in both fundamental and applied directions. On the fundamental side, the likelihood framework opens avenues for implementing model variants on spatial cues specifications (e.g. frequency-dependent spectral weighting~\cite{llado_spectral_2025}), or dynamic scenarios~\cite{llado_predicting_2024,barumerli_frambi_2025}. On the applied side, the likelihood warrants consideration as a perceptual evaluation metric for HRTFs in machine learning pipelines. The present results demonstrate that it successfully differentiates interpolation methods by their perceptual impact on localisation behaviour (Sec.~\ref{ssec:comparison}), capturing individual differences that standard acoustic measures cannot resolve. Log-spectral distortion, for instance, treats all spatial directions equally and has been shown not to reliably predict perceptual localisation quality~\cite{yao_perceptually_2024} nor colouration~\cite{mckenzie_toward_2025}. Because the likelihood integrates over the full response distribution, weighting spatial regions by their actual contribution to directional estimates, it offers a behaviourally grounded alternative that does not require an explicit mapping from acoustic to perceptual features: a general formulation of which remains an open problem~\cite{fantini_survey_2025, geronazzo_strong_2025}.

\section{Conclusions}
This work presents two contributions to computational auditory modelling for human sound localisation. On the methodological side, we formulated an explicit likelihood for the Barumerli et al. (2023)~\cite{barumerli_bayesian_2023} model and validated it through a complete Bayesian workflow~\cite{barumerli_frambi_2025}. The two-stage fitting procedure reliably identifies sensorimotor variability from as few as 99 trials, while the moderate recovery of spectral and prior parameters exposes structural identifiability limits. On the applied side, the interpolation comparison showed that full-sphere coverage and spectral fidelity determine template quality, while the specific algorithm is secondary once these conditions are met. More broadly, the validated likelihood extends in both fundamental and applied directions: as a foundation for statistical auditory modelling and as a perceptually grounded evaluation metric for HRTFs in machine learning pipelines. The open-source Python implementation and reproducible validation workflow are released to facilitate the adoption of these methods across the spatial hearing community.
\acknowtext
\vspace{-3mm}
We thank Rapolas Daugintis, Katarina C. Poole, and Ludovic Pirard for providing behavioural data that enabled this analysis.

Claude (Anthropic) was used as a writing assistant for language editing and clarity; all content was reviewed and approved by the authors, who bear full responsibility for the final text.
\vspace{-3mm}

\funding
\vspace{-3mm}
This work was supported by the European Union's Horizon Europe research and innovation programme under the Marie Skłodowska-Curie grant agreement No. 101201118 (project MIA, awarded to R.B.), and the Horizon 2020 research and innovation funding programme under grant agreement No. 101017743 (project SONICOM).
\vspace{-3mm}
\conflict
\vspace{-3mm}
The authors declare no competing financial interests.
\vspace{-7mm}
\dataavailability
\vspace{-3mm}
The Python implementation of the model will be made publicly available upon publication at \href{https://github.com/robaru/bayesian_listener}{\texttt{github.com/robaru/bayesian\_listener}}. 
Analysis notebooks will be available at \href{https://github.com/robaru/bayesian_listener_notebooks}{\texttt{github.com/robaru/bayesian\_listener\_notebooks}}. 
Behavioural data from participants are available at the \href{https://ecosystem.sonicom.eu/databases/58}{SONICOM data ecosystem}. 
The measured HRTF dataset is available at the \href{https://transfer.ic.ac.uk:9090/#/2022_SONICOM-HRTF-DATASET}{Imperial College data transfer service}.
\vspace{-5mm}
\authorcontrib
\vspace{-3mm}
R.B. conceived and designed the study, developed the initial Python implementation, conducted analyses, and drafted the manuscript. F.B. contributed to interpolation methodology, Python implementation, and manuscript writing. E.Z. and A.H. contributed to porting the model from MATLAB to Python and Python package development. M.G. and L.P. provided critical manuscript revision.

\bibliographystyle{ieeetr}
\bibliography{references}
%
%
\newpage
\supplementary
\beginsupplement
\section{Template interpolation distortion}
\label{supsec:temp_dist}
The three panels illustrate the spectral distortion introduced by different template construction approaches for the G.R.A.S.\ KEMAR dummy head from the SONICOM database~\cite{engel_sonicom_2023}. The unprocessed HRTF templates show the expected frequency-dependent amplitude structure across elevation, with clear spectral cues in the 4--16\,kHz range known to support elevation perception. The \texttt{barumerli2023} method returns spectral amplitudes only at measured HRTF directions, preserving this structure above the horizon but leaving elevations below $-45^\circ$ unrepresented, as these are absent from the measurement grid (green line). Extending the method to a full spherical grid (\texttt{barumerli2023 - Full grid}) recovers coverage at all elevations but introduces severe spectral distortions below the measurement boundary, where extrapolation becomes unreliable.

\begin{figure}[h]
    \centering
    \includegraphics[width=.8\linewidth]{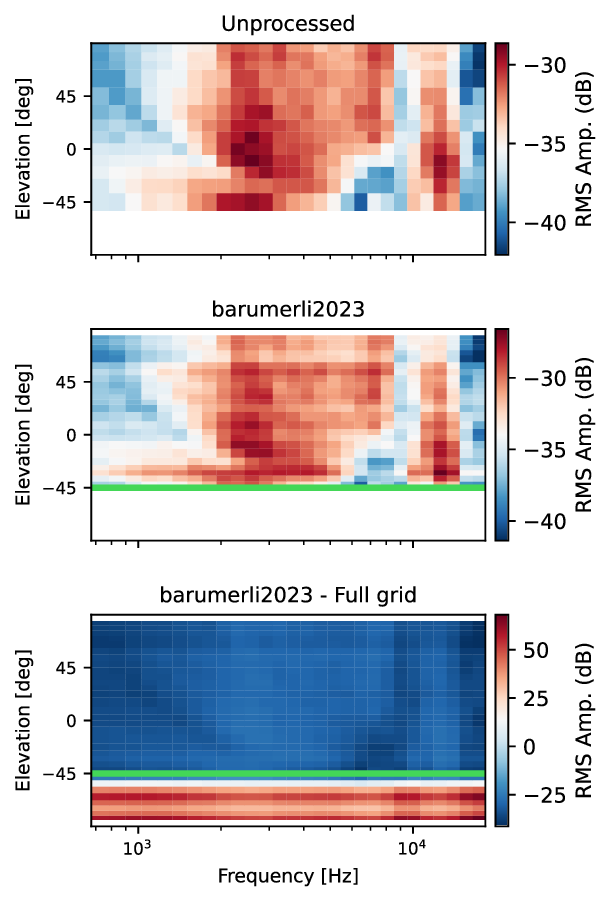}
    \caption{Template spectral amplitudes as a function of elevation and frequency for the G.R.A.S.\ KEMAR dummy head for the median plane. Top: unprocessed HRTF spectral amplitude profiles. Middle: templates reconstructed using the \texttt{barumerli2023} method on a quasi-uniform spherical grid, limited to elevations above the minimum available in the HRTF measurement grid as implemented in the original model (green line)~\cite{barumerli_bayesian_2023}. Bottom: the same method extended to a full spherical grid, revealing severe amplitude distortions below the measurement boundary.}
    \label{fig:sh_supplementary}
\end{figure}


\section{Likelihood stability}
\label{supsec:lik_stab}
\begin{figure}[h]
    \centering
    \includegraphics[width=.8\linewidth]{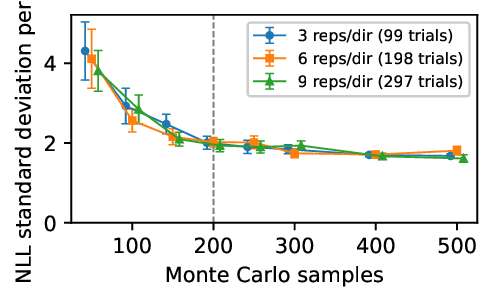}
    \caption{Per-trial negative log-likelihood (NLL) standard deviation as a function of Monte Carlo samples, averaged over 28 parameter configurations (mean~$\pm$ standard deviation). The dashed vertical line marks the selected value of 200 samples.}
    \label{fig:NLL_variability}
\end{figure}

The number of Monte Carlo samples required for stable likelihood approximation was determined by computing the per-trial negative log-likelihood (NLL) standard deviation across the 28 ground-truth parameter configurations (Section~4) as a function of sample count. Figure~\ref{fig:NLL_variability} shows that variability decreases rapidly up to 200 samples, with diminishing returns beyond this point, which was therefore selected for all subsequent analyses.
\section{Parameter recovery}
\label{supsec:par_rec}
We defined 28 parameter combinations spanning the expected range of listener variability ($\sigma_\mathrm{mon}$: $2$--$15$, log-spaced; $\sigma_\mathrm{prior}$: $5^\circ$--$90^\circ$, log-spaced; $\sigma_\mathrm{m}$: $5^\circ$--$16^\circ$, linearly spaced), with pairwise correlations between ground-truth parameters kept below $|r| < 0.1$ to ensure that recovery performance could be assessed independently for each parameter. For each combination, we simulated localisation responses using a KEMAR HRTF set processed with SHMAX interpolation, presenting 33 target directions with 3 repetitions each (99 trials), matching the minimum trial count in the experimental data.

\begin{figure}[h]
    \centering
    \includegraphics[width=.7\linewidth]{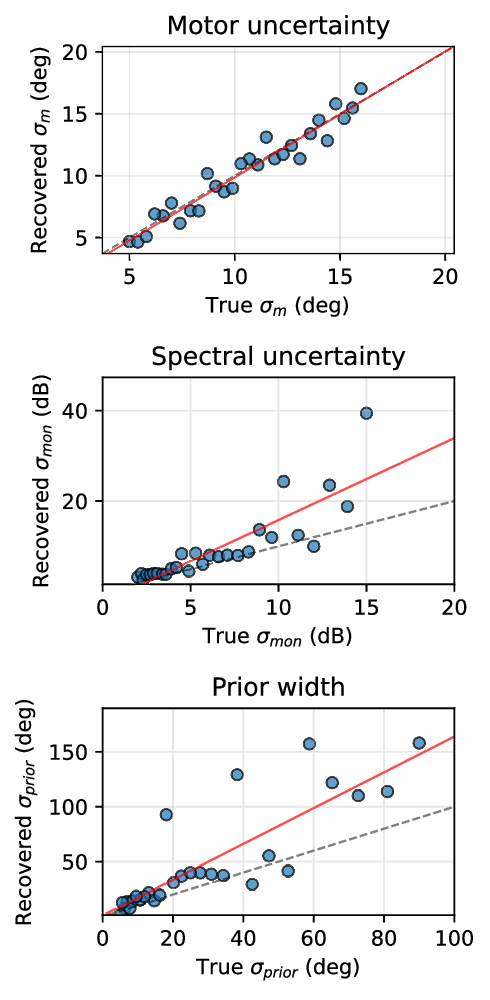}
    \caption{Parameter recovery results for the free model parameters. Each panel compares ground-truth (x-axis) and recovered (y-axis) values across the 28 synthetic parameter configurations. The dashed grey line indicates perfect recovery (identity); the solid red line is the Pearson regression fit. Recovery quality is quantified by Pearson $r$ and mean bias (recovered minus ground-truth), reported in the main text.}
    \label{fig:recovery}
\end{figure}

Figure~\ref{fig:recovery} shows ground-truth against recovered parameter values for each of the three free parameters. 

\end{document}